\documentstyle[11pt,aaspp4,cuifig]{article}
\slugcomment{To appear in The Astrophysical Journal Letters}

\begin{document}
\title{Discovery of High-Frequency QPOs in Black Hole Candidate 
XTE J1859+226}
\author{Wei Cui\altaffilmark{1}, 
Chris R. Shrader\altaffilmark{2}\altaffilmark{3}, 
Carole A. Haswell\altaffilmark{4}, and 
Robert I. Hynes\altaffilmark{5}}

\altaffiltext{1}{Center for Space Research, Massachusetts
Institute of Technology, Cambridge, MA 02139; cui@space.mit.edu}
\altaffiltext{2}{Laboratory for High-Energy Astrophysics, NASA/Goddard 
Space Flight Center, Greenbelt, MD 20771; shrader@grossc.gsfc.nasa.gov}
\altaffiltext{3}{also Universities Space Research Association}
\altaffiltext{4}{Dept of Physics and Astronomy, The Open University, 
Walton Hall, Milton Keynes, MK7 6AA, United Kingdom;
C.A.Haswell@open.ac.uk}
\altaffiltext{5}{Dept of Physics and Astronomy, University of
Southampton, Southampton, SO17 1BJ, United Kingdom; 
rih@astro.soton.ac.uk}

\begin{abstract}
We report the discovery of quasi-periodic oscillations (QPOs) at 
roughly 187 Hz and 150 Hz in the X-ray intensity of X-ray nova
XTE J1859+226. The source was observed during a recent outburst 
with RXTE. Besides these high-frequency QPOs, we have also detected 
QPOs (and sometimes their harmonics) at 6--7 Hz, and significant 
broad-band variability at low frequencies. These properties, as 
well as the observed hard X-ray spectrum, make XTE J1859+226 a 
black hole candidate (BHC). 

The detection of QPOs at two distinct frequencies $\gtrsim 100$ Hz 
in two observations separated by about 4 hours provide additional 
insights into the high-frequency QPO phenomenon in BHCs. The importance 
lies in the proposed interpretations which invariably involve the 
effects of strong gravity near a black hole. We compare our results 
to those of other BHCs, and discuss the impact of the observational 
data on the models in a global context.
\end{abstract}


\section{Introduction}
A new X-ray nova, designated as XTE J1859+226, was discovered by the
ASM/RXTE on October 9, 1999 (Wood et al. 1999). Subsequent observations 
(2-60 keV) of the source with the PCA/RXTE revealed a hard X-ray 
spectrum roughly of power-law 
shape (Markwardt et al. 1999). The source was detected up to 200 keV 
(McCollough \& Wilson 1999; Dal Fiume et al. 1999). It is interesting 
to note that the hard X-ray flux of the source (as measured by 
the BATSE/CGRO; McCollough \& Wilson 1999) began to decrease while 
the soft X-ray flux (as measured by the ASM) was still rising. This 
is remarkably similar to another X-ray nova XTE J1550-564 (Cui et al. 
1999) and may be common for such objects. 

Model fits to the X-ray spectrum of XTE J1859+226 implied a column 
density about 3--8 $\times 10^{21}$ $cm^{-2}$ along the line of sight 
(Markwardt et al. 1999; Dal Fiume et al. 1999). The relatively low 
foreground extinction facilitated rapid optical follow-up 
observations. The optical counterpart (Garnavich et al. 1999), as 
well as the radio counterpart (Pooley \& Hjellming 1999), was 
identified soon after the discovery. The optical spectrum 
of XTE J1859+226 showed characteristics typical of soft X-ray 
transients during an X-ray outburst (Wagner,~R.~M. et al. 1999; 
Hynes et al. 1999). The optical/UV continuum spectrum from the 
HST/STIS observations imply a binary period $<$ 1 day (Hynes et
al. 1999). Subsequent optical observations indeed detected a possible 
period of 9.15 hours (Garnavich \& Quinn 2000), which are likely
associated with the orbital motion.

XTE J1859+226 showed a strong quasi-periodic oscillation (QPO) in 
X-ray intensity during the rising phase of the outburst. The QPO was 
first detected at 0.45 Hz (Markwardt et al. 1999) and 
moved up to $\sim$5.5 Hz as the outburst proceeded (Dal Fiume et al. 
1999), similar to other X-ray novae (Fox \& Lewin
1998; Cui et al. 1998, 1999; Dieters et al. 2000). The combination 
of the hard X-ray spectrum and QPOs makes XTE J1859+226 a black hole 
candidate (BHC). Here we report the detection of QPOs at much higher 
frequencies.

\section{Observations}
Fig.~1 shows the ASM light curve of XTE J1859+226. The outburst 
consists of periods of fast rise and
exponential decay, which used to be considered typical for X-ray
novae (Chen et al. 1997). The rise time (10\%--90\% peak)
is roughly 5 days and the decay time (e-fold) about 23 days. The 
source experienced short-lived flares near the peak of the outburst,
and a secondary maximum (at MJD 51515--51540) during the decay 
phase. The source continued to fade, following the secondary 
maximum, with roughly the same e-fold time. A much expanded light
curve, shown in the inset, focuses around the peak of the
outburst and clearly reveals short X-ray flares.

We obtained data from four pointed PCA observations of XTE J1859+226 
around the time of the outburst peak (see Table 1). The first three 
observations consist of one satellite orbit and the fourth of two 
orbits. The count rate of the source varies by less than 30\% between
these observations and the X-ray spectrum remains roughly the same. 
Note that only a subset of the five detector 
units (a.k.a. PCUs) were turned on during the observations and the 
number of PCUs that were on varied.

\section{Data Analysis and Results}
We only used data from the high-resolution {\em Single-Bit} and 
{\em Event} modes.
We first rebinned the {\em Event} data to the same resolution as the
{\em Single-Bit} data (i.e., $2^{-13}$ s) and then combined both sets.
For the analyses, we defined two energy bands: 2--5.9 keV (soft, from 
the low-energy {\em Single-Bit} data) and 5.9--60 keV (hard, from the 
combination of the high-energy {\em Single-Bit} data and the {\em Event} 
data). For each energy band, 
we performed a fast Fourier transform for every 128 s data 
segment of each observation. We normalized the power density spectrum
(PDS) following the procedure of Leahy et al. (1983). Each PDS was 
then properly weighted and all the spectra co-added to obtain the 
average PDS for that observation in the chosen energy band. Fig.~2 
shows the results for the hard band. 

Below about 1 Hz, the PDS is dominated by broad-band variability. The 
spectrum of such noise is roughly of 1/f shape, although it is clearly 
more complicated in Obs. 3. Between 1--20 Hz, QPOs are
prominent. Except for Obs. 3, where only one QPO is detected, 
the QPOs appear to be harmonically related, 
with the stronger peak being the fundamental component. There 
is a hint of a sub-harmonic component, especially in Obs. 1. 
The QPO is quite narrowly peaked but is 
much broader in Obs. 3. To quantify the QPO properties, we fit 
each PDS with an empirical model consisting of a power law (for the 
broad-band noise), Lorentzian functions (for the QPOs), and a constant 
(for Poisson noise). We limited the 
fit to the frequency range 0.01--20 Hz, except for Obs. 3, 
where we chose the 0.3--20 Hz range (since the broad-band noise is 
roughly of power-law shape only above 0.3 Hz; see Fig.~2). Table~2 
shows the results. As expected, the two QPOs are 
harmonically related for observations 1, 2, and 4. The sub-harmonic 
component is significantly detected only in Obs. 1.

At frequencies above 20 Hz, broad but localized features
are present in the PDS for observations 1, 2, and 4, but not for 
Obs. 3 (see the insets in Fig.~2). The features found in 
observations 1 and 4 are clearly QPO like, although fairly broad,
and similar to the high-frequency QPOs found in GRO J1655-40 
(Remillard et al. 1999a), XTE J1550-564 (Remillard et al. 1999b), 
and 4U 1630-47 (Remillard \& Morgan 2000). Again, we fit the 
high-frequency portion 
(20--4000 Hz) of each PDS with the same empirical model as described 
above. Note that near 20 Hz the PDS is still significantly affected by 
the first harmonic of the low-frequency QPO whenever it is present, so 
the power-law component of the model actually mimics the
high-frequency wing of the QPO harmonic. The best-fit parameters are
also shown in Table~2. 

A broad QPO is significantly detected at roughly 187 and 150 Hz in 
observations 1 and 4, respectively. The Q ($\equiv f/\Delta f$) of the 
signal is small, on the order of unity. Only a broad ``shoulder'' 
($Q \approx 0.5$) appears to be present in Obs. 2, and no 
feature is significantly detected in Obs. 3. In the latter 
case, however, the $3 \sigma$ upper limit on the fractional rms 
amplitude is $\sim$2.7\% for the 187 Hz QPO and $\sim$3.1\% for the 
150 Hz QPO, which are not very constraining due to poor 
statistics. To explore any energy dependence of such QPOs, we
performed independent searches for a QPO in the soft and hard bands. 
The search in the soft band failed to yield any significant 
detections. As an example, Fig.~3 shows the PDS from Obs. 1 for 
each band. Only after fixing the Lorentzian frequency and width to 
those of the QPO detected in the hard band, we obtained, from the 
fit, an rms amplitude of $1.0^{+0.4}_{-0.8}$\% for the signal in the
soft band. Therefore, the QPO is definitely stronger in the hard 
band; so is the 150 Hz QPO detected in Obs. 4.

For completeness, we also constructed an average cross power spectrum 
between the soft and hard bands for each observation (following a
procedure similar to that described in, e.g., Cui 1999b), in order to 
measure possible 
phase lags. The results (also shown in Table 2) reveal significant 
{\em hard lag} (i.e., the X-ray emission in the hard band lags behind 
that in the soft band) associated with the fundamental component of 
the low-frequency QPO for observations 1, 2, and 4, but not for 
Obs. 3, while the broad-band variability shows {\em soft lags} 
in the vicinity of the QPO. The measured QPO lag (i.e., the difference 
between columns 4 and 5 in Table 2) is as large as $\sim$0.55 radians, 
which corresponds to a time lag $\sim$15 ms.

\section{Discussion}
The most important result from this work is the detection of QPOs at
high frequencies ($\gtrsim 100$ Hz) in XTE J1859+226. These QPOs most 
probably originate from a 
single periodic process that varies in its period. It is remarkable 
that the period of the process can change on hour timescales. Our 
detections are corroborated by Markwardt et al. (2000), who found 
QPOs at $\sim$174 Hz from their own observations. Very similar 
behavior is seen in XTE J1550-564: a QPO was first discovered at 
$\sim$184 Hz during a giant X-ray flare near the peak of the outburst 
(Remillard et al. 1999b) and another was later detected at $\sim$284 
Hz at a much lower flux (Homan, Wijnands, \& van der Klis 1999). 
These QPOs are, however, in sharp contrast to the high-frequency QPOs 
seen in microquasars, GRS 1915+105 (Morgan, Remillard, \& Greiner 1997) 
and GRO J1655-40 (Remillard et al. 1999a), which were thought to 
maintain a constant frequency. This implies that there could be
intrinsic physical differences underlying such QPOs in BHCs. 
Alternatively, the different behaviors may merely result from 
insufficient statistics or sporadic coverage. For XTE J1859+226, the 
QPOs are stronger at higher energies, in terms of their fractional rms
amplitudes, which is typical of nearly all QPOs in BHCs (review by Cui 
1999a). 

Several models have been proposed to explain the high-frequency QPOs
in BHCs, all of which invoke strong gravity near
the central black hole (review by Cui, Chen, \& Zhang 1998). It was
first suggested that the QPOs might be associated with the Keplerian
motion at the last stable orbit around Schwarzschild black holes 
(Morgan, Remillard, \& Greiner 1997). Perhaps the motion of ``blobs''
or ``hot spots'' in an accretion disk extending all the way to
the last stable orbit somehow causes the modulation of X-ray emission. 
However, the emissivity of the disk vanishes at the
last stable orbit, due to the torque-free boundary condition, so it 
is perhaps more sensible to associate the QPOs with orbiting hot 
spots located in a region where the disk emissivity peaks. While 
appealing for its simplicity the model is incompatible with the 
spectral results
(Cui, Chen, \& Zhang 1998). Nowak et al. (1997) subsequently proposed 
that the QPOs are associated with the oscillation modes in the
accretion disk. Not only can such modes 
be supported by the disk, they also become trapped in the innermost 
portion of the disk, purely due to relativistic effects (Kato \& 
Fukue 1980; Nowak \& Wagoner 1992, 1993). Finally, Cui et al. (1998) 
argued that the QPOs could be due to the Lense-Thirring precession 
of the tilted circular orbits of blobs or hot spots at the inner 
edge of the accretion disk, motivated by observational evidence for 
the presence of rapidly rotating black holes in microquasars (Zhang, 
Cui, \& Chen 1997). 

All the models were initially proposed to explain the ``constant
frequency QPOs'' in microquasars. Consequently, the accretion disk 
was assumed to extend to the last stable orbit in the models. Since 
the high-frequency QPOs now appear to be common among BHCs, an obvious 
question to ask now is whether the models can accommodate the QPOs 
that vary in frequency, as in the case of XTE J1550-564 and 
XTE J1859+226. For those models that involve Keplerian motion or 
Lense-Thirring precession, ``variable frequency QPOs'' can be
explained if the 
inner edge of the accretion disk is allowed to move. Because the 
precession frequency in the Lense-Thirring model depends strongly 
on the radius of the orbit (roughly, $f \propto r^{-3}$), 
a $\sim$35\% change in the QPO frequency (as in 
the extreme case of XTE J1550-564) requires that the inner 
edge of the disk moves radially outward only by $\sim$15\%. Therefore, 
the variable frequency QPOs would pose least serious challenges to 
this model. In comparison, the dependence 
of the Keplerian frequency on the orbital radius is not as steep; 
the inner edge of the disk would be required to move about twice 
as far in the Keplerian motion model. The problem is far more 
serious (if not fatal) for the disk oscillation model, because the 
oscillation modes tend to be fairly well localized in the accretion 
disk. While much improved statistics are clearly needed to ultimately 
establish the physical origin of the high-frequency QPOs in BHCs, it 
is clear that the discovery of variable frequency QPOs has begun to 
seriously constrain the models.

An additional similarity between XTE J1859+226 and XTE J1550-564 is the
presence of a prominent and evolving QPO at a few Hz during the rising
phase of the outburst. However, the QPO is almost certainly of a 
different physical origin for the two sources, because the phase 
lag associated with the fundamental component is of an {\em opposite} 
sign (cf. Cui, Zhang, \& Chen 2000). While the complicated QPO lags 
in XTE J1550-564 are likely the same phenomenon as was first
discovered in a subset of the QPOs of GRS 1915+105 (Cui 1999b), the 
hard QPO lag in XTE J1859+226 makes the QPO here similar to a second 
type of the low-frequency QPOs in 
XTE J1550-564, which were detected at a later stage of the outburst 
and also showed a sub-harmonic component (Wijnands, Homan, \& 
van der Klis 1999). The phase lag measurement has, therefore, added 
new puzzles to the already very intriguing QPO phenomenology in BHCs
(see Cui 1999b for a discussion of possible implications of different 
QPO lags on the models).

It has been argued, based on the observed dependence of QPO properties 
on mass accretion rate (as indicated by X-ray flux) and photon energy, 
that the QPOs in BHCs form a heterogeneous class of phenomena, each of 
which may well be produced by a different physical mechanism (Cui 1999a; 
Cui et al. 1998). For a given source, the QPOs are known to come and 
go in an unpredictable manner, and a different set of QPOs may
dominate the PDS at different times (see an extreme example offered 
by GRS 1915+105; Morgan, Remillard, \& Greiner 1997). Here we have
shown that such changes took place on time scales as short as roughly 
one hour in XTE J1859+226 (i.e., the time between Obs. 3 and 
Obs. 2 or 4). The source clearly behaved very differently 
during Obs. 3, both in 
terms of the QPOs and broad-band noise (see Fig.~2). It is interesting 
that the timing properties (low- and high-frequency QPOs plus 
broad-band variability) all seem to be correlated. A plausible 
model must, therefore, be able to account for such a global change.  

\acknowledgments
We thank Jean Swank and Craig Markwardt for useful discussions. W.~C. 
acknowledges financial support of NASA grants NAG5-7484 and
NAG5-7990. C.~A.~H. and R.~I.~H were supported by grant F/00-180/A 
from the Leverhulme Trust.

\clearpage
\begin{deluxetable}{ccccccc}
\tablecolumns{7}
\tablewidth{0pc}
\tablecaption{PCA Observation Log\tablenotemark{1}}
\tablehead{
\colhead{Obs.}&\colhead{Obs. Id.}&\colhead{Obs. Time (UT)\tablenotemark{2}}&\colhead{
PCUs} &\colhead{Exp. (s)} &\colhead{CR (c/s)} &\colhead{HR}}
\startdata
1 & 01-03 & 02:04:00--03:12:00 & 4 & 2640 & 3028 & 0.46 \nl
2 & 01-02 & 03:40:00--04:47:00 & 4 & 3060 & 3104 & 0.45 \nl
3 & 01-01 & 05:16:00--06:23:00 & 3 & 3480 & 2406 & 0.41 \nl
4 & 01-00 & 06:52:00--10:01:00 & 3 & 7440 & 2446 & 0.44 \nl
\tablenotetext{1}{Columns: (1) designated observation number, (2)
observation ID with prefix ``40122--01--'' omitted, (3) observation
time interval, (4) number of PCUs on, (5) exposure time, (6) mean count
rate per PCU, and (7) mean hardness ratio (the mean count rate of the 
5.9--60 keV band divided by that of the 2--5.9 keV band).}
\tablenotetext{2}{All observations were carried out on 18 October
1999.}
\enddata
\end{deluxetable}

\clearpage

\begin{deluxetable}{ccccccccc}
\footnotesize
\tablecolumns{9}
\tablewidth{0pc}
\tablecaption{QPO Properties\tablenotemark{1}}
\tablehead{
 &\multicolumn{5}{c}{Low-Frequency QPO\tablenotemark{2}} &
\multicolumn{3}{c}{High-Frequency QPO\tablenotemark{3}} \\
\cline{2-6} \cline{7-9} \\
\colhead{Obs.}&\colhead{Freq.}&\colhead{FWHM}&\colhead{RMS}
&\colhead{Peak Lag\tablenotemark{4}}&\colhead{Cont. Lag\tablenotemark{5}}
&\colhead{Freq.}&\colhead{FWHM}&\colhead{RMS} \\
 &\colhead{(Hz)}&\colhead{(Hz)}&\colhead{(\%)}&\colhead{(radian)}
&\colhead{(radian)}&\colhead{(Hz)}&\colhead{(Hz)}&\colhead{(\%)} 
}
\startdata
1&$5.97^{+0.01}_{-0.01}$&$0.44^{+0.02}_{-0.02}$&$7.38^{+0.09}_{-0.08}$&$0.31^{+0.04}_{-0.04}$&$-0.24^{+0.04}_{-0.04}$&$187^{+14}_{-11}$&$95^{+46}_{-43}$&$3.3^{+0.5}_{-0.7}$\nl 
 &$2.92^{+0.08}_{-0.09}$&$0.40^{+0.18}_{-0.18}$&$0.8^{+0.1}_{-0.1}$& &
& \nl
 &$11.74^{+0.05}_{-0.06}$&$2.1^{+0.2}_{-0.2}$&$3.12^{+0.10}_{-0.09}$&
& & \nl
2&$5.97^{+0.01}_{-0.01}$&$0.54^{+0.02}_{-0.02}$&$6.50^{+0.08}_{-0.08}$&$0.23^{+0.05}_{-0.05}$&$-0.24^{+0.06}_{-0.06}$&$82^{+56}_{-26}$&$180^{+58}_{-37}$&$3.8^{+0.3}_{-0.4}$\nl
 &$11.6^{+0.09}_{-0.09}$&$2.6^{+0.3}_{-0.3}$&$2.9^{+0.1}_{-0.1}$& & & \nl
3&$7.3^{+0.3}_{-0.2}$&$5.8^{+1.0}_{-0.9}$&$3.3^{+0.3}_{-0.3}$&$-0.24^{+0.13}_{-0.13}$&$-0.26^{+0.12}_{-0.12}$&\nodata&\nodata&\nodata \nl
4&$5.95^{+0.01}_{-0.01}$&$0.57^{+0.02}_{-0.02}$&$5.31^{+0.06}_{-0.06}$&$0.24^{+0.04}_{-0.04}$&$-0.15^{+0.07}_{-0.07}$&$150^{+17}_{-28}$&$124^{+53}_{-39}$&$3.2^{+0.5}_{-0.6}$\nl
 &$10.9^{+0.2}_{-0.3}$&$4.5^{+0.7}_{-0.6}$&$2.9^{+0.2}_{-0.2}$& & & \nl
\tablenotetext{1}{The uncertainties shown represent roughly $1 \sigma$
errors. Note that the fractional rms amplitudes are for the 5.9--60 keV 
band.}
\tablenotetext{2}{Including (sub-)harmonics.}
\tablenotetext{3}{For Obs. 2, the signal has very low Q, more
like a broad ``shoulder'' than a QPO (see text). }
\tablenotetext{4}{The peak lag is computed by averaging the measured
phase lag over a frequency interval ($f_0 - w/5$, $f_0 + w/5$), where 
$f_0$ is the centroid frequency of the QPO and $w$ the width (FWHM).}
\tablenotetext{5}{The continuum lag is computed by averaging the 
measure phase lag over (1, 10) Hz with ($f_0 - 5 w$, $f_0 + 5 w$) 
excluded, except for Obs. 3 where the lag is averaged over (1, 15) Hz 
excluding ($f_0 - w$, $f_0 + w$). Note that the difference between the
peak lag and the continuum lag gives the phase lag associated with the
QPO. } 
\enddata
\end{deluxetable}

\clearpage
\begin{figure}
\psfig{figure=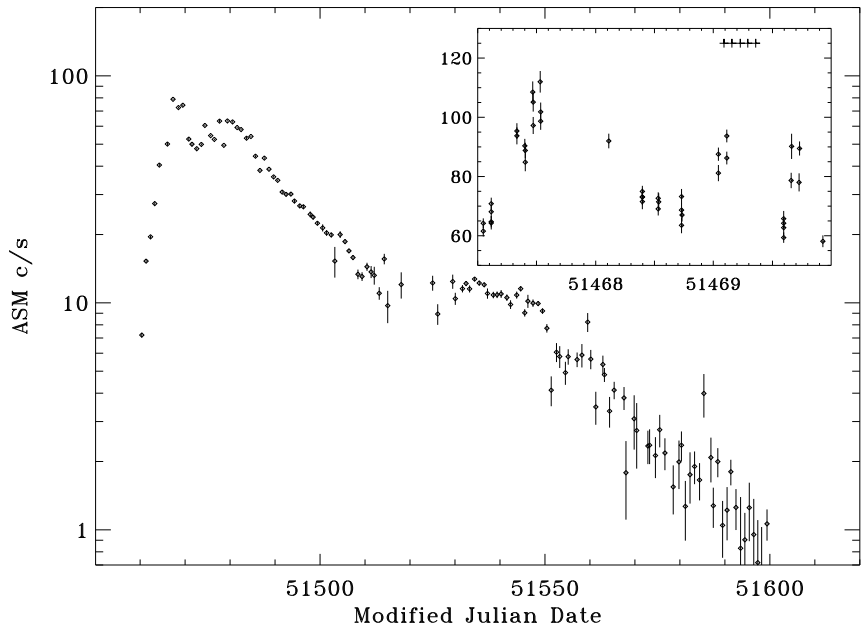,width=5in,angle=90}
\caption{Daily-averaged ASM light curve of XTE J1859+226 during a
recent X-ray outburst. The inset expands the light curve around the
peak of the outburst when the PCA observations were carried out. 
Note that we chose to plot 90 dwell data in the inset for better 
coverage of the brief flares. Each cross marks the start time of 
a ``good time interval'' for each satellite orbit. For reference, 
MJD 51450 corresponds to 1999 September 29. }
\end{figure}

\begin{figure}
\psfig{figure=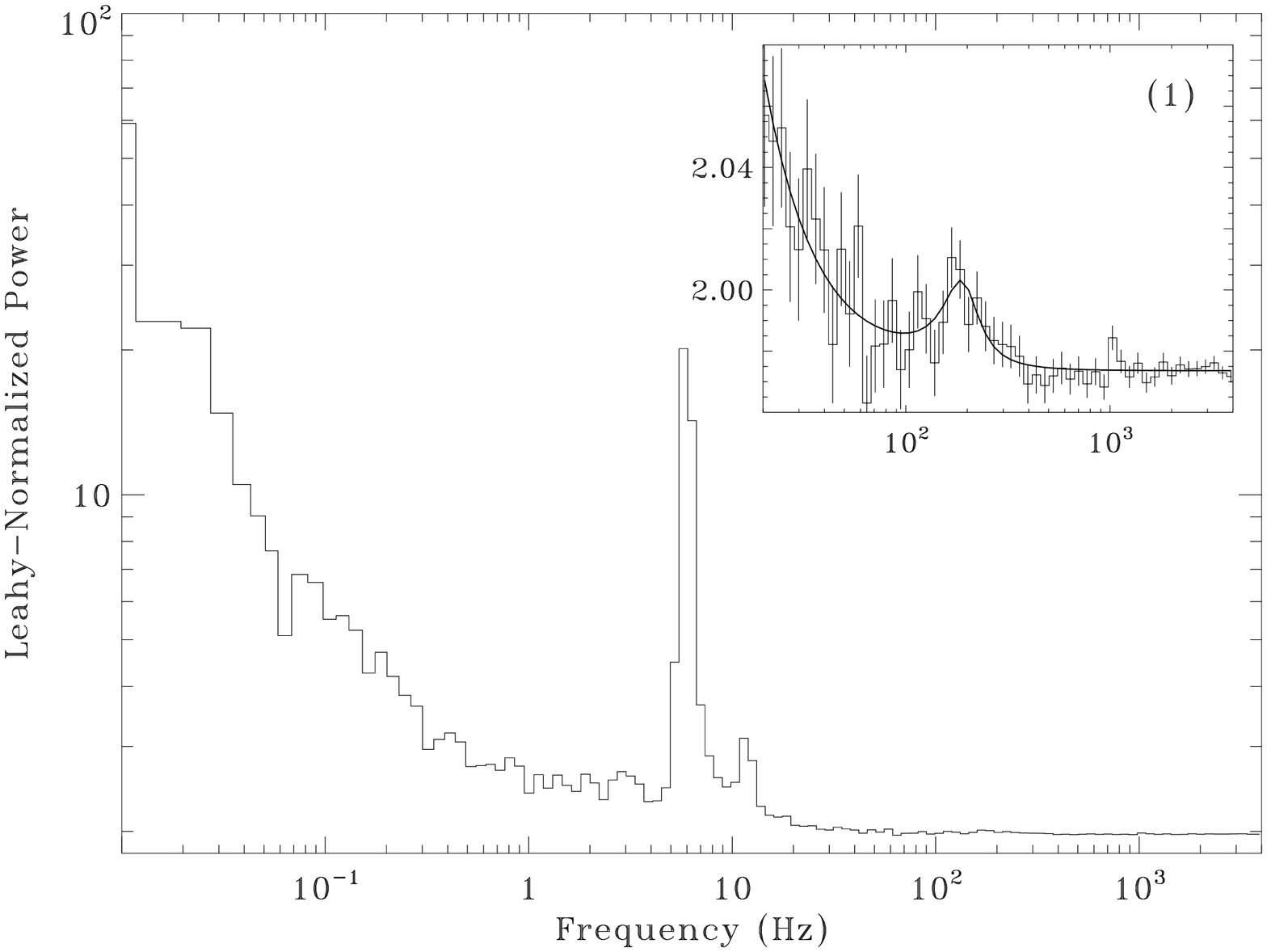,width=3.2in,angle=90}
\psfig{figure=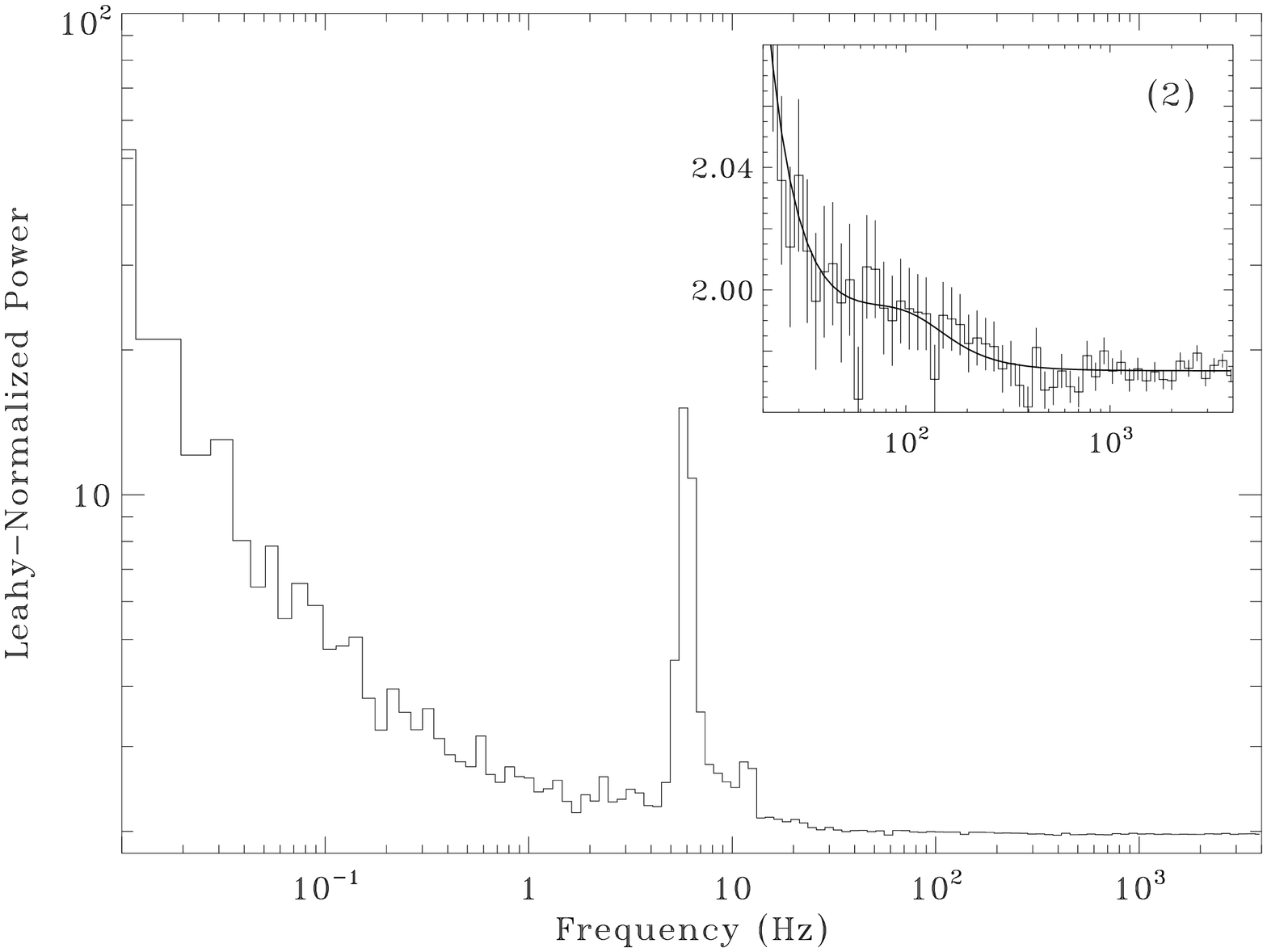,width=3.2in,angle=90}
\psfig{figure=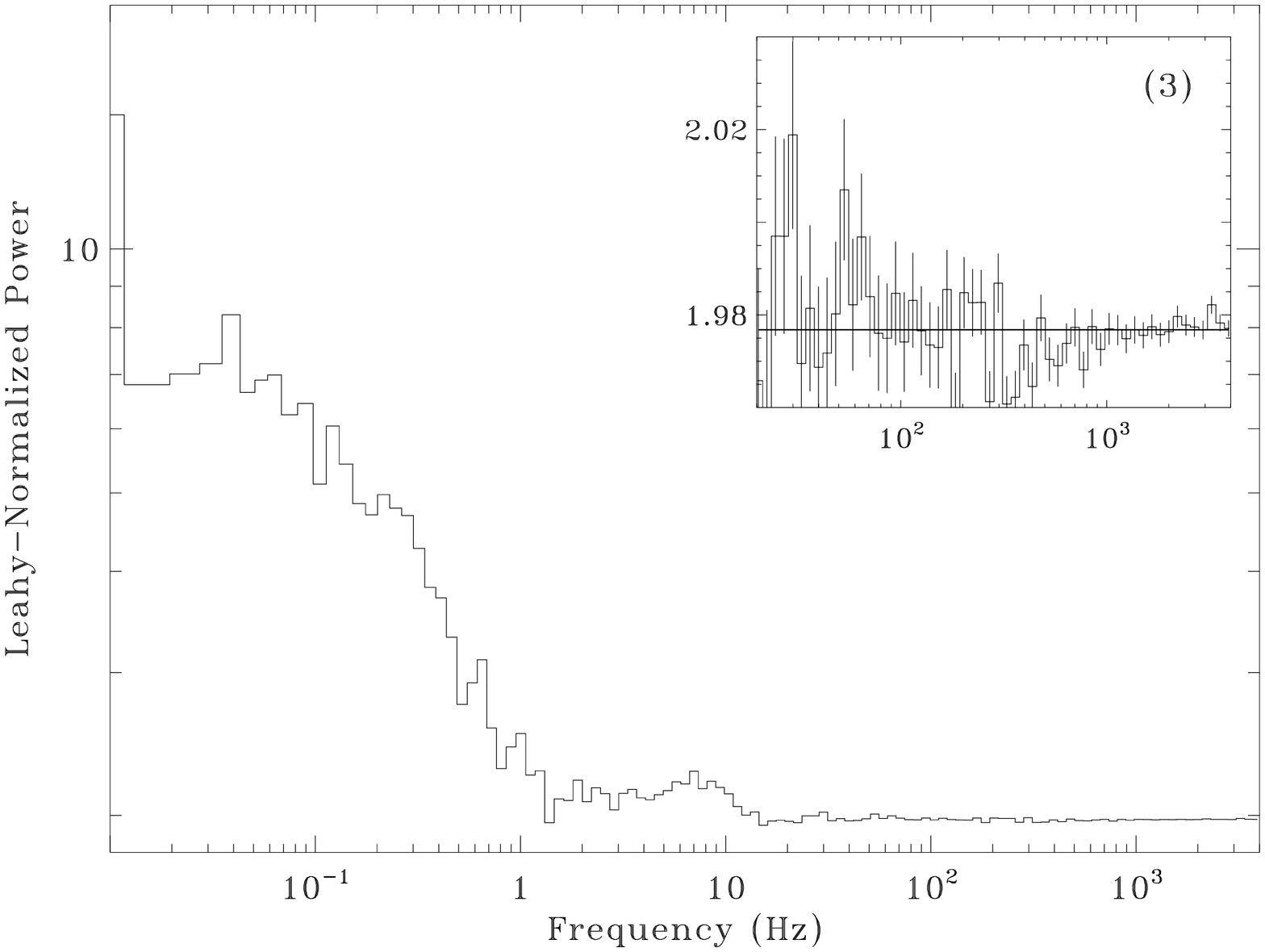,width=3.2in,angle=90}
\psfig{figure=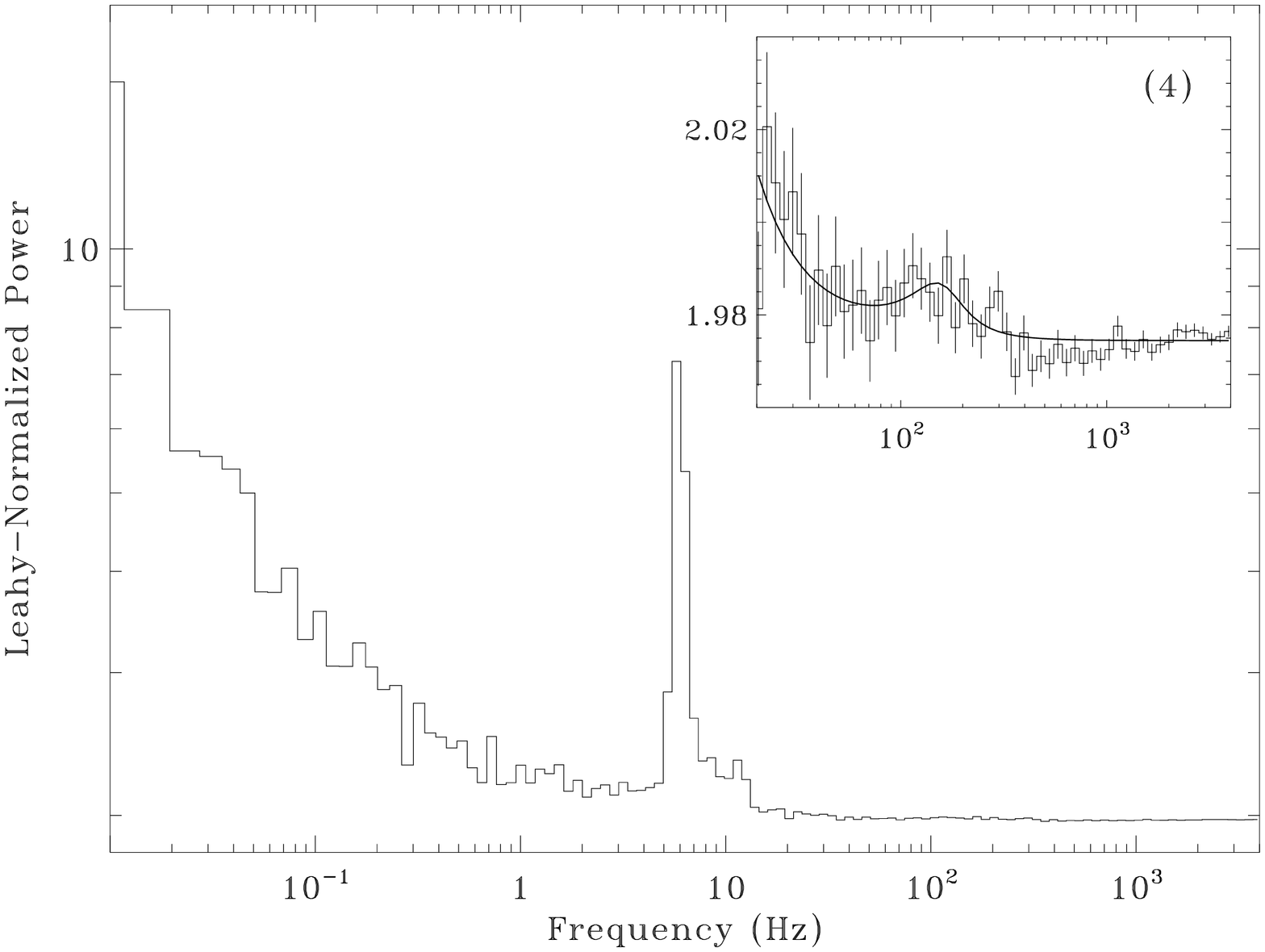,width=3.2in,angle=90}
\caption{Power density spectra of XTE J1859+226. The insets provide an
expanded view of the frequency range where the high-frequency QPOs are
found. Note that the solid lines represent the best-fit to the data. }
\end{figure}

\begin{figure}
\psfig{figure=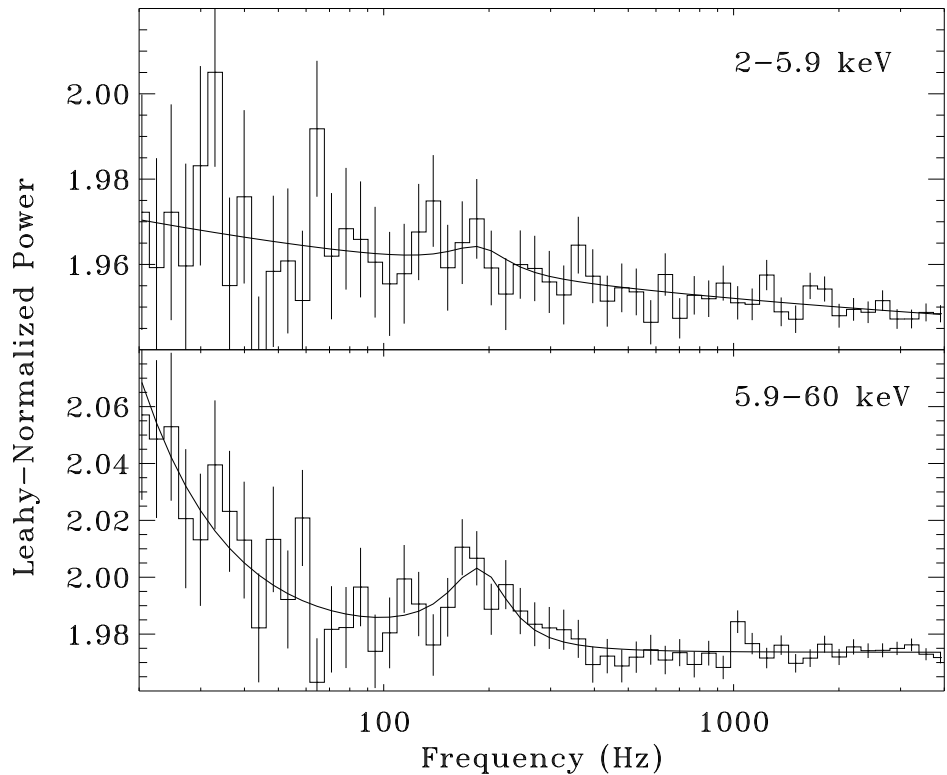,width=5in}
\caption{Power density spectra of XTE J1859+226 in two energy bands. 
These are made from observation 1, and zoom onto the frequency range 
where the high-frequency QPO is detected. The solid lines represent 
the best-fit model to the data, although for the soft band the
frequency and width of the Lorentzian was fixed to those derived from 
the hard band (see text). }

\end{figure}

\end{document}